\documentclass[runningheads]{llncs}
\usepackage[T1]{fontenc}
\usepackage{graphicx}

\usepackage{amsmath}
\usepackage{amsfonts}
\usepackage{amssymb}
\usepackage{xcolor}

\usepackage{hyperref} 

\usepackage[caption=false,subrefformat=parens,labelformat=parens]{subfig}

\usepackage{float}

\usepackage{enumitem}

\newcommand{\evUniv}{\mathcal{U}_{ev}}
\newcommand{\objUniv}{\mathcal{U}_{obj}}
\newcommand{\objTypeUniv}{\mathcal{U}_{otype}}
\newcommand{\evTypeUniv}{\mathcal{U}_{etype}}

\newcommand{\attrUniv}{\mathcal{U}_{attr}}
\newcommand{\valUniv}{\mathcal{U}_{val}}
\newcommand{\qualUniv}{\mathcal{U}_{qual}}
\newcommand{\timeUniv}{\mathbb{T}}

\newcommand{\stringUniv}{\mathcal{U}_{\Sigma}}

\newcommand{\varNamesUniv}{\mathcal{U}_{\mathrm{obVar}}}
\newcommand{\objVarUniv}{\mathcal{U}_{\mathrm{obVar}}}

\newcommand{\nodeNameUniv}{\mathcal{U}_{\mathrm{evVar}}}
\newcommand{\evVarUniv}{\mathcal{U}_{\mathrm{evVar}}}

\newcommand{\bindingSetNames}{\mathcal{U}_{\mathrm{setName}}}

\newcommand{\constrFuncStandalon}{\ensuremath{constr}}
\newcommand{\constrFunc}[1]{\ensuremath{\constrFuncStandalon(#1)}}
\newcommand{\predSet}{\mathbb{P}_L}

\usepackage{eurosym}

\usepackage{soul}
\setul{0.5ex}{0.15ex}

\def\mi#1{\mathit{#1}}

\RequirePackage{tikz}
\usetikzlibrary{calc, shapes, arrows, positioning, patterns, decorations.pathreplacing, backgrounds, petri, automata,arrows, trees, graphs}

\newcommand{\dom}{\mathrm{dom}}

\usepackage{zed-csp}

\usepackage{mathtools}

\usepackage{xargs}
\usepackage[misc,geometry]{ifsym}

\begin{document}
\title{OCPQ: Object-Centric Process \\Querying \& Constraints}
%
%
\author{Aaron Küsters\,\textsuperscript{\Letter}\,\orcidID{0009-0006-9195-5380} \and
Wil~M.P.~van~der~Aalst\orcidID{0000-0002-0955-6940}}
\authorrunning{A.~Küsters and W.M.P.~van~der~Aalst}
%
\institute{Chair of Process and Data Science (PADS), RWTH Aachen University
\\
\email{\{kuesters,wvdaalst\}@pads.rwth-aachen.de}}
\setcounter{tocdepth}{1}
\maketitle              
\begin{abstract} 
Process querying is used to extract information and insights from process execution data.
Similarly, process constraints can be checked against input data, yielding information on which process instances violate them.
Traditionally, such process mining techniques use case-centric event data as input.
However, with the uptake of Object-Centric Process Mining (OCPM), existing querying and constraint checking techniques are no longer applicable.
Object-Centric Event Data (OCED) removes the requirement to pick a single case notion (i.e., requiring that events belong to exactly one case) and can thus represent many real-life processes much more accurately.
In this paper, we present a novel highly-expressive approach for object-centric process querying, called \emph{OCPQ}.
It supports a wide variety of applications, including OCED-based constraint checking and filtering.
The visual representation of nested queries in OCPQ allows users to intuitively read and create queries and constraints.
We implemented our approach using (1) a high-performance execution engine backend and (2) an easy-to-use editor frontend.
Additionally, we evaluated our approach on a real-life dataset, showing the lack in expressiveness of prior work and runtime performance significantly better than the general querying solutions SQLite and Neo4j, as well as comparable to the performance-focused DuckDB.

\end{abstract}
\section{Introduction}
\label{sec:intro}
In organizations, process execution data contain valuable insights that are often not leveraged to their full extent.
The domain of process querying is concerned with methods and techniques for extracting such insights from event data.
For example, given data of an order management process, a simple query for interesting cases could be formulated in natural language as ``Find all cases where \texttt{pay order} is executed more than once''.
Process querying of execution data also has a strong correspondence to process constraints:
Identifying violations of a process constraint, e.g., ``\texttt{pay order} should be executed exactly once per case'', corresponds to querying its violations (i.e., a query with the negated constraint).

To promote using this opportunity of gaining insights, it is important to allow stakeholders to query interesting scenarios in their processes themselves.
For that reason, graphical notations for constraints or queries are often introduced, as a way to also allow stakeholders without programming experience to utilize them.
Similarly, there are also graphical declarative process models, like DECLARE~\cite{pesicDECLAREFullSupport2007}, in which visual models describe an underlying set of constraint rules.

More recently, process querying and constraint approaches based on \emph{Object-Centric Event Data} (OCED) have been proposed~\cite{aalstObjectCentricBehavioralConstraints2017,esserMultiDimEventData2021,parkMonitoringConstraintsBusiness2022}.
OCED no longer assumes a single case notion in data, i.e., that events belong to exactly one defined case.
Instead, OCED contains a set of objects and a set of events of specified types.
OCED also allows for relationships between objects and events, as well as between objects and objects. 
As such, OCED can represent many real-life processes much more accurately than traditional, flat event logs.
For example, in an order management process, the different objects interacting in the process could include \texttt{customers}, \texttt{orders}, \texttt{items}, \texttt{packages}, and \texttt{employees}.
Therefore, classical case-centric approaches, which assign one object per event, are too limiting~\cite{aalstObjectCentricPMUnraveling2023}.
These advantages of OCED translate directly to process querying techniques based on OCED.
In particular, a more accurate and interconnected representation of the underlying real-life process can be queried, instead of only a flat representation that could lead to inaccurate results or misleading conclusions.

\begin{figure}[h]
    \centering
    \includegraphics[trim={0 0 1.7cm 0.3cm},clip,width=0.75\textwidth]{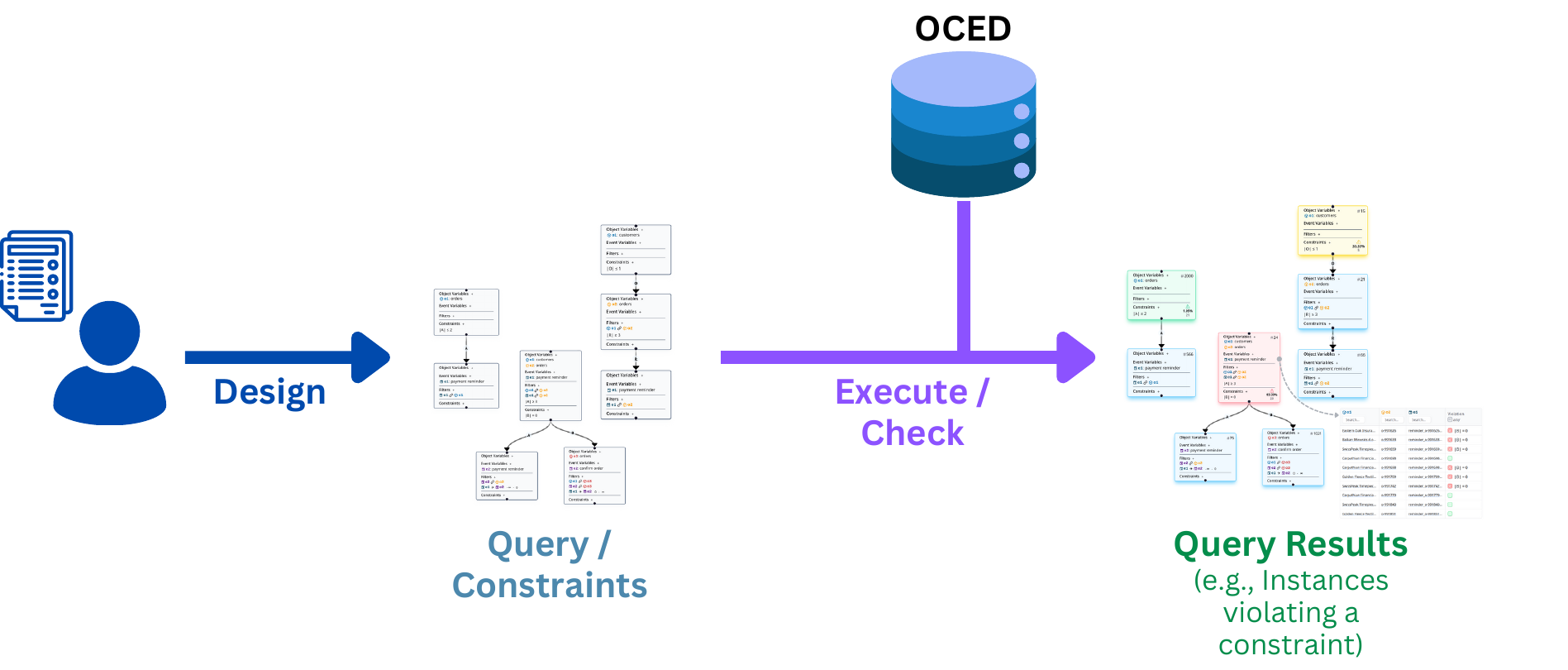}
    \caption{Overview of the object-centric process querying and constraint approach.}
    \label{fig:intro-fig}
\end{figure}

In this paper, we present an object-centric nested querying approach with an accompanying full graphical tool implementation, focusing on high expressiveness, fast runtime performance, and easy usability.
An overview of the approach, including inputs and outputs, is shown in \autoref{fig:intro-fig}.
First, stakeholders design queries or constraints, optionally based on some regulation or specification document.
The created queries can then be evaluated based on input OCED, yielding query results.
The query results are visually shown in aggregation (e.g., with the total number of results or the percentage of violating instances) but can also be explored in the tool individually or exported (e.g., to a CSV or XLSX file).

Consider the following example of an object-centric constraint for an order management process:
``If a third payment reminder for an order is sent to a customer, no (other) order by this customer should be confirmed afterwards''.
What at first seems to be a rather simple rule actually involves multiple different object and event types, as well as multiple instances of the same object type (i.e., multiple order objects).
In fact, to the best of our knowledge, none of the so far proposed graphical process querying or constraint techniques can express such a rule.
Even in all-purpose querying languages, like SQL or Cypher, expressing such a rule is not only unapproachable for people without programming skills, but also largely impractical for larger data because of their poor performance for such types of queries.

\begin{figure}[h]
    \centering
    \includegraphics[width=0.9\textwidth]{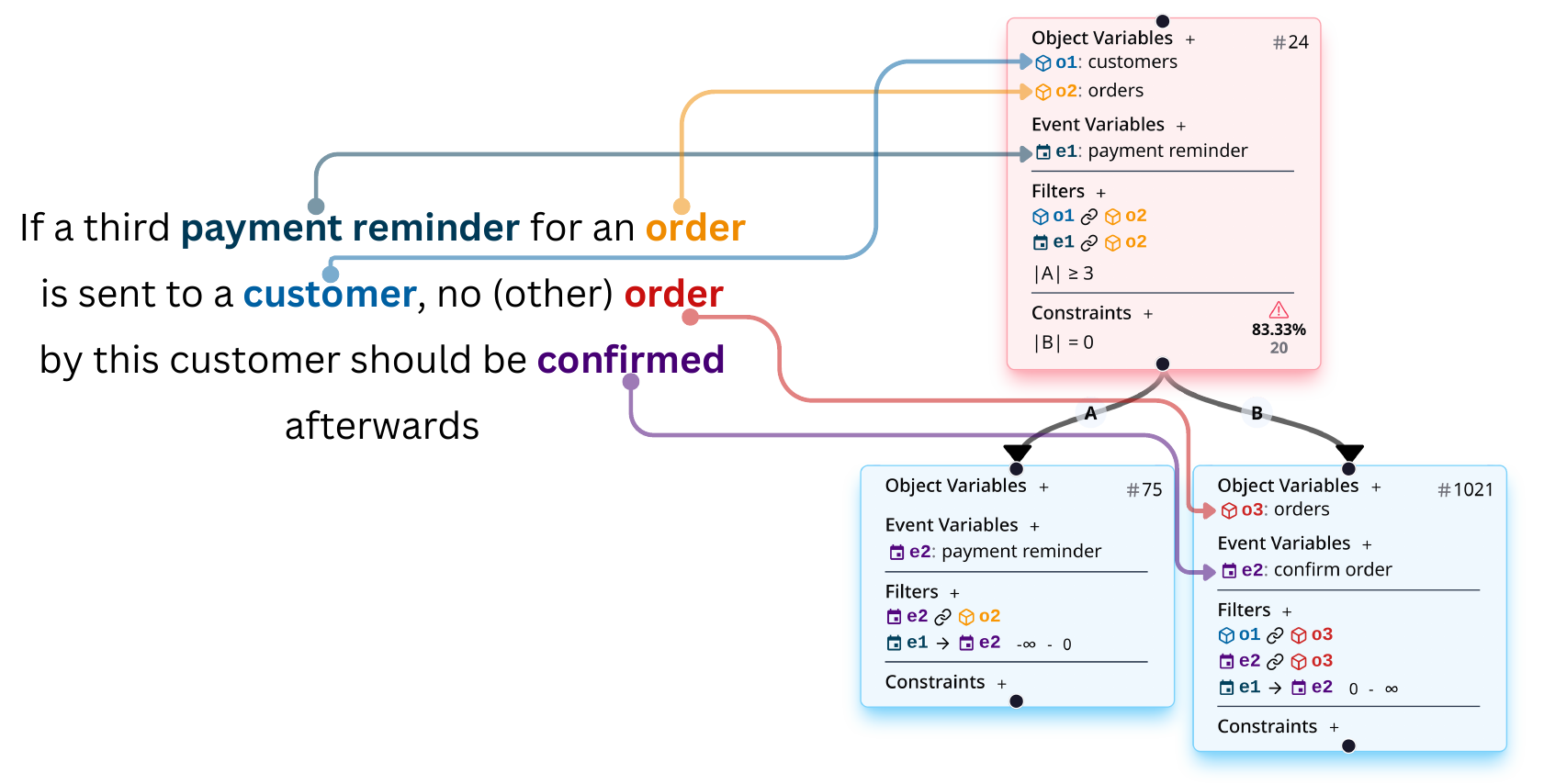}
    \hspace{0.5cm}
    \caption{The mapping of an example constraint in natural language to a visual nested query constraint in OCPQ.
    The included arrows indicate how objects or events mentioned in the textual description are modeled in the visual constraint.
    }
    \label{fig:example-constraint-translation}
\end{figure}

We fill this gap by introducing \emph{OCPQ}, an object-centric process querying approach, that leverages the full flexibility of the OCED data model, while focusing on easy, visual usability and a very fast runtime performance.
We also present a full graphical tool implementation of the approach, which is publicly available at \url{https://github.com/aarkue/ocpq}.
In contrast to related work, OCPQ can also express advanced queries and constraints spanning multiple object and event types while still allowing for easy visual modeling.
For instance, the previously introduced example can be modeled visually as a nested query constraint in OCPQ, as shown in \autoref{fig:example-constraint-translation}.

In our tool implementation, nested queries are evaluated through a recursive, parallelizable algorithm, achieving good runtime performance and demonstrating the feasibility of our approach, even for larger, real-life datasets.

The remainder of this paper is structured as follows.
In \autoref{sec:related}, we first discuss related work.
Next, we introduce preliminaries in \autoref{sec:prelims}.
\autoref{sec:querying} describes the main concepts of our approach.
Our tool implementation is covered in \autoref{sec:impl}, followed by an evaluation of the expressiveness and runtime in \autoref{sec:eval}.
Finally, we conclude this paper in \autoref{sec:conclusion}.

\section{Related Work}
\label{sec:related}
In this section, we present related work on process querying, process constraints and declarative process models.

Process querying research covers filtering and manipulation of process repositories, which can contain process models as well as process executions (i.e., event data), based on a (formal) query.
In this paper, we focus on querying of event data without corresponding process models.
In~\cite{perez-alvarezProcessInstanceQuery2022}, the authors describe the \emph{Process Instance Query Language} (PIQL), which allows querying the number of cases or events fulfilling specified criteria.
The proprietary \emph{Celonis Process Querying Language} (Celonis PQL) introduced in \cite{vogelgesangCelonisPQLQuery2022} is another domain-specific process querying language heavily inspired by SQL.
Through the integration of Celonis PQL with the underlying data model schema, table joins do not need to be specified explicitly, as would be the case with \texttt{JOIN} statements in SQL.
All the previously mentioned process querying research is primarily focused on traditional, flat event data.
However, in~\cite{esserMultiDimEventData2021}, Esser et al.~describe storing multidimensional (i.e., object-centric) event data in graph databases and using the universal graph querying language \emph{Cypher} of the \emph{Neo4j} database system to query entities or subgraphs.

Apart from querying, process constraints as well as declarative process models are also important related fields for this paper.
Both of these fields are concerned with checking if input event data satisfies specified rules, and returning violating fragments of data if not.
Most notably, \emph{DECLARE}~\cite{pesicDECLAREFullSupport2007} was the first declarative approach for business process management.
While internally, DECLARE uses Linear Temporal Logic (LTL), parametrized LTL templates are represented visually (e.g., as specific types of arrows) to ease usability.
An initial constraint template language, also called DECLARE, is included by default, but custom template languages can be created and used as well.

In~\cite{schonigEfficientCustomisableDeclarative2016}, Schöning et al.~present an SQL-based approach for discovering declarative process constraints.
By creating subqueries for returning activity combinations or computing support and confidence values, the discovery of constraints that satisfy given thresholds can be implemented as a simple SQL \texttt{SELECT} query.
The authors report generally competitive performance metrics for discovering constraints.
However, expanding the set of discoverable constraints, for example to ternary instead of only binary constructs, would lead to significant higher execution times.
In~\cite{schoningDiscoveryOfMultiPerspectiveDeclarative2016}, the authors extend this approach to also consider data attributes, resource, and time perspectives.
Similarly, in~\cite{rivaSQLBasedDeclarativeProcess2023}, the authors present a declarative process mining framework based on SQL queries, allowing for discovery and checking of constraints.
The authors additionally evaluated the time required for executing the resulting query sets across different database schemas.

There are also a few object-centric process constraint approaches.
OCBC models, introduced in \cite{aalstObjectCentricBehavioralConstraints2017}, combine behavioral constraints, inspired by DECLARE patterns, with object class data models imposing cardinality rules.
As such, OCBC models can express that relationships between objects on the class-level should also manifest on the behavior-level (e.g., only items belonging to a purchase order should be associated with its pick item events).
In \cite{parkMonitoringConstraintsBusiness2022}, the authors introduce \emph{Object-Centric Constraint Graphs} (OCCGs).
These constraint graphs can capture interactions between objects and events, as well as control-flow between event types, based on a given object type.
Additionally, performance metrics regarding events can be included.

\section{Preliminaries}
\label{sec:prelims}
As preliminaries, we first define some basic mathematical concepts.

For a set $X$, the \emph{powerset} of $X$ is $\mathcal{P}(X) = \{Y \mid Y \subseteq X  \}$.
We also use partial functions: Given two sets $A$ and $B$, a partial function $f\colon A \not \rightarrow B$ maps \emph{some} of the elements of $A$ to values in $B$.
For elements $x$ that are not mapped to a value (i.e., $x \not \in \dom(f)$), we write $f(x) = \bot$.
If two partial functions have disjoint domains (i.e., $\dom(f) \cap \dom(g) = \emptyset$), we write $f \union g$ for their symmetric union.
Moreover, we use subset notation (i.e., $f \subseteq g$) if it holds that $\forall_{x \in \dom(f)}\; g(x) = f(x)$.

Next, we define the universes that form the basis of our formalization.
\begin{definition}
	Let $\stringUniv$ be the universe of strings.
	We use the following pairwise disjoint universes:
	\begin{itemize}[nosep]
		\item \makebox[2.2cm][l]{$\evUniv \subseteq \stringUniv$} Universe of events \hfill(e.g., $e_1$)
		\item \makebox[2.2cm][l]{$\objUniv \subseteq \stringUniv$} Universe of objects \hfill(e.g., $o_1$)
		\item \makebox[2.2cm][l]{$\evTypeUniv \subseteq \stringUniv$} Universe of event types (\emph{activities}) \hfill(e.g., $\texttt{confirm order}$)
		\item \makebox[2.2cm][l]{$\objTypeUniv \subseteq \stringUniv$} Universe of object types \hfill(e.g., $\texttt{orders}$)
		\item \makebox[2.2cm][l]{$\attrUniv \subseteq \stringUniv$} Universe of attribute names \hfill(e.g., $\texttt{time}$)
		\item \makebox[2.2cm][l]{$\qualUniv \subseteq \stringUniv$} Universe of relationship qualifiers \hfill(e.g., $\texttt{places}$)
		\item \makebox[2.2cm][l]{$\objVarUniv \subseteq \stringUniv$} Universe of object variable names \hfill(e.g., $\texttt{o1}$)
		\item \makebox[2.2cm][l]{$\nodeNameUniv \subseteq \stringUniv$} Universe of event variable names \hfill(e.g., $\texttt{e1}$)
		\item \makebox[2.2cm][l]{$\bindingSetNames \subseteq \stringUniv$} Universe of set variable names \hfill(e.g., $\texttt{A}$)
	\end{itemize}
	We write $\mathbb{T}$ for all possible timestamps and durations and $\valUniv$ for the universe of all attribute values (with, for instance, $\mathbb{T}\subseteq \valUniv$ and $\stringUniv  \subseteq \valUniv$).
\end{definition}

\begin{figure}[h]\scriptsize
		\renewcommand{\arraystretch}{0.87}
		\subfloat[$\mi{oaval}$ with a visualization of the O2O relationships.]
		{
		\begin{tabular}{c|c|c|c}
			Object & Attribute          & Timestamp                               & Value                                                  \\ \hline
			$o_1$  & $\texttt{type}$    & $\ast$                                  & \texttt{customers}                                     \\
			$o_1$  & $\texttt{objects}$ & $\ast$                                  & $\{(\texttt{places},o_2)\}$                            \\
			$o_1$  & $\texttt{city}$     & \scriptsize \texttt{2016-01-06T14:15} & \texttt{Bonn}                                            \\
			$o_1$  & $\texttt{city}$     & \scriptsize \texttt{2018-09-03T10:32} & \texttt{Aachen}                                            \\
			$o_2$  & $\texttt{type}$    & $\ast$                                  & \texttt{orders}                                        \\
			$o_2$  & $\texttt{objects}$ & $\ast$                                  & $\{(\texttt{\scriptsize contains},o_3), (\texttt{\scriptsize contains},o_4)\}$ \\
			$o_3$  & $\texttt{type}$    & $\ast$                                  & \texttt{items}                                         \\
			$o_4$  & $\texttt{type}$    & $\ast$                                  & \texttt{items}                                         \\
		\end{tabular}
		\hspace{1cm}
		\begin{minipage}[c]{2cm}
			\includegraphics[width=2cm]{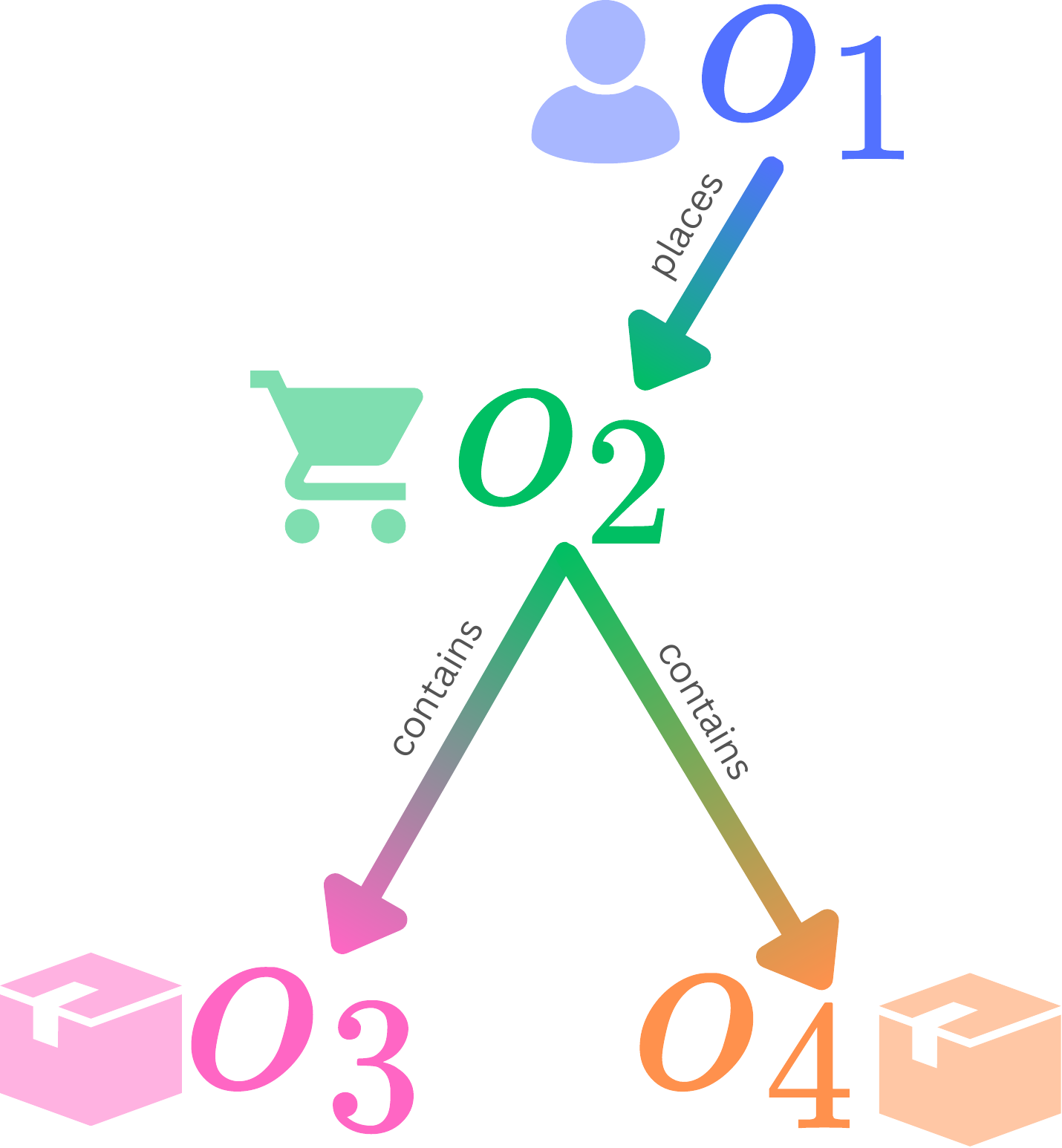}
		\end{minipage}
\centering
\label{tab:example-oced-objects}
		}

		\vspace*{-0.2cm}
		\subfloat[$\mi{eaval}$ with a visualization of the O2O and E2O relationships.]
		{\centering
		\begin{tabular}{c|c|c}
			Event & Attribute           & Value                                                                                       \\ \hline
			$e_1$ & $\texttt{activity}$ & \texttt{place order}                                                                        \\
			$e_1$ & $\texttt{objects}$  & $\{(\texttt{customer},o_1), (\texttt{order},o_2),$ \\
			 &  & \hfill $(\texttt{item},o_3),(\texttt{item},o_4)\}$ \\
			$e_2$ & $\texttt{activity}$ & \texttt{pack item}                                                                          \\
			$e_2$ & $\texttt{objects}$  & $\{(\texttt{item},o_3)\}$                                                                   \\
			$e_3$ & $\texttt{activity}$ & \texttt{pack item}                                                                          \\
			$e_3$ & $\texttt{objects}$  & $\{(\texttt{item},o_4)\}$                                                                   \\
			$e_4$ & $\texttt{activity}$ & \texttt{ship items}                                                                         \\
			$e_4$ & $\texttt{objects}$  & $\{(\texttt{ships},o_3),(\texttt{ships},o_4)\}$                                             \\
			$e_5$ & $\texttt{activity}$ & \texttt{payment reminder}                                                                   \\
			$e_5$ & $\texttt{objects}$  & $\{(\texttt{recipient},o_1),(\texttt{order},o_2)\}$                                         \\
			$e_5$ & $\texttt{fee}$      & 15                                                                                         \\
			$e_6$ & $\texttt{activity}$ & \texttt{pay order}                                                                          \\
			$e_6$ & $\texttt{objects}$  & $\{(\texttt{order},o_2)\}$                                                                  \\
		\end{tabular}
		\hspace{1cm}
		\begin{minipage}[c]{4cm}
			\includegraphics[width=4cm]{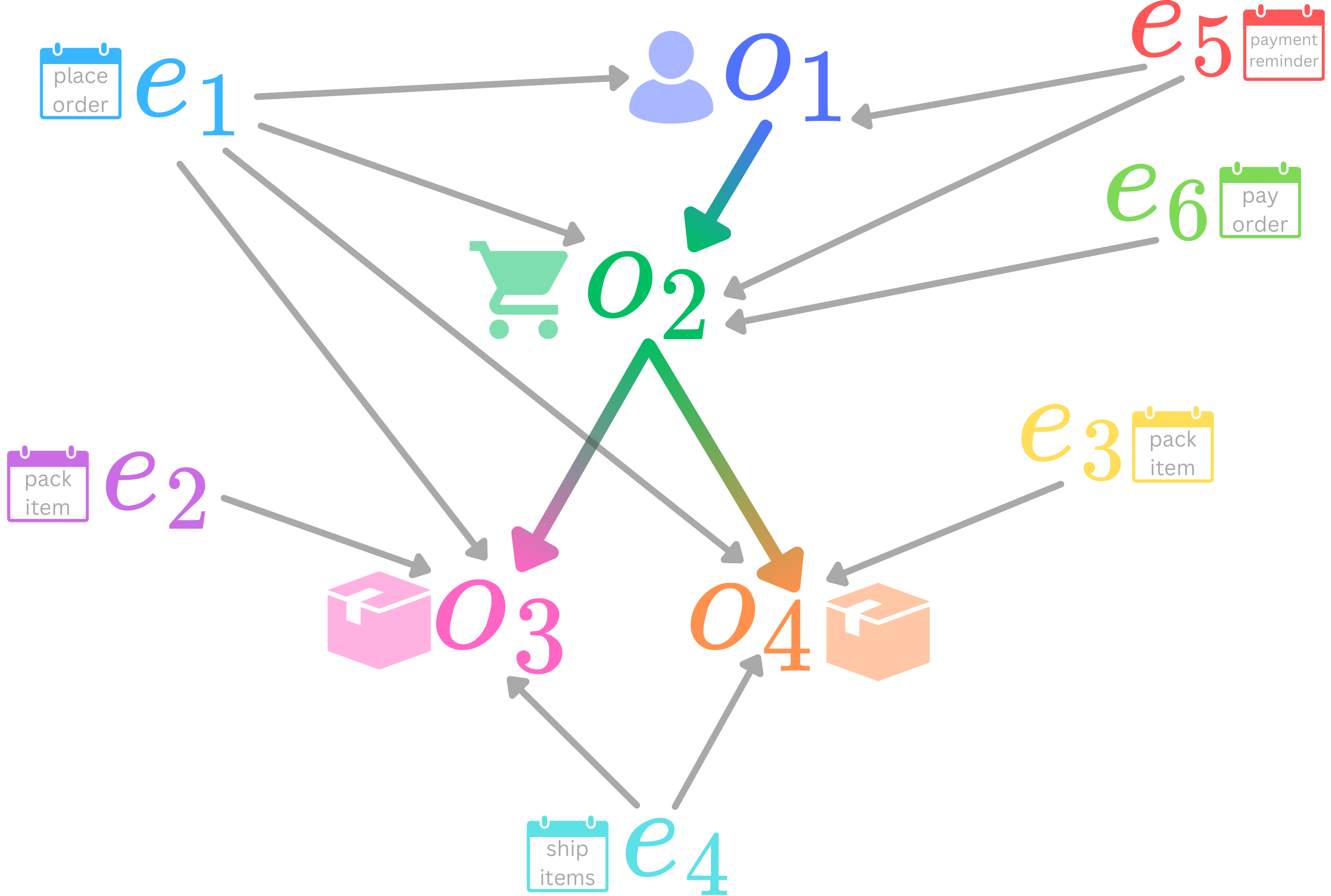}
		\end{minipage}
		\label{tab:example-oced-events}}
		\centering
		\caption{Example $\mi{oaval}$ and $\mi{eaval}$ tables with corresponding relationship visualizations.}
		\label{tab:example-oced-general}
	\end{figure}

Next, we introduce Object-Centric Event Data (OCED) formally.
Our definition is inspired by the OCEL~2.0 specification\footnote{\url{https://www.ocel-standard.org/}}, but the presented approach is not limited to any particular OCED model.
At the very core, OCED contains a set of objects and a set of events, each of which have additional attributes.
For objects, these attributes can also change over time.
Certain types of attributes (e.g., for the object type or relationships between objects) are mandatory.
\begin{definition}
	Object-Centric Event Data (OCED) can be described as a tuple $L=(E,O,\mi{eaval},\mi{oaval})$ of the following components:
	\begin{itemize}[nosep]
		\item \textbf{Events} $E \subseteq \evUniv$ as the set of events.
		\item \textbf{Objects} $O \subseteq \objUniv$ as the set of objects.
		\item \textbf{Event Attributes} $\mi{eaval}: E \rightarrow (\attrUniv \not \rightarrow \valUniv)$, which provides attribute values for events.
		      For convenience, we write $\mi{eaval}_{e} = \mi{eaval}(e)$ for an $e \in E$ as a shorthand.
		      The following properties have to hold for $\mi{eaval}$:
		      \begin{itemize}
			      \item $\forall_{e\in E}\; \mi{eaval}_e(\texttt{activity}) \in \evTypeUniv$: each event has exactly one \emph{event type}.
			      \item $\forall_{e\in E}\;\mi{eaval}_e(\texttt{objects}) \subseteq \qualUniv \times O \land  \mi{eaval}_e(\texttt{objects}) \neq \emptyset$: each event has at least one qualified reference to an object.
			      \item $\forall_{e\in E}\; \mi{eaval}_e(\texttt{time}) \in \timeUniv$: each event has a timestamp.
		      \end{itemize}
		\item \textbf{Object Attributes} $\mi{oaval}: O \rightarrow (\attrUniv \times \timeUniv \not \rightarrow \valUniv)$, which provides the attribute values of an object $o\in O$ at a concrete timestamp.
		      For convenience, we write $\mi{oaval}_o^t(attr) = \mi{oaval}(o)(attr,t)$ for a given $o \in O, t\in \timeUniv$ and $attr \in \attrUniv$ as a shorthand.
		      The following properties have to hold for $\mi{oaval}$:
		      \begin{itemize}
				\item For the \emph{time-stable} attributes $a \in \{\texttt{objects}, \texttt{type}\} \subseteq \attrUniv$, the assigned value must not change over time. In particular, it should hold that $\forall_{o\in O}\forall_{t \in \timeUniv}\forall_{t' \in \timeUniv}\; \mi{oaval}_o^t(a) = \mi{oaval}_o^{t'}(a)$.
				We also write $\mi{oaval}_o(a)= \mi{oaval}_o^t(a,t)$ with an arbitrary timestamp $t\in \timeUniv$ for these attributes.
			      \item $\forall_{o\in O}\; \mi{oaval}_o(\texttt{type})\in \objTypeUniv$: every object has exactly one \emph{object type}.
			      \item $\forall_{o\in O}\; \mi{oaval}_o(\texttt{objects})\subseteq \qualUniv \times O$: an object can, optionally, contain qualified references to (other) objects.
		      \end{itemize}
	\end{itemize}
\end{definition}
As an example OCED, consider $L=(E,O,\mi{eaval},\mi{oaval})$ with a set of objects $O = \{o_1, o_2, o_3, o_4\}$ and a set of events $E = \{ e_1, e_2, e_3, e_4, e_5, e_6\}$.
	The attribute values of all objects and events, assigned by $\mi{eaval}$ and $\mi{oaval}$, are shown in \autoref{tab:example-oced-general}.
	For $\mi{oaval}$, time-stable object attributes are marked with an $\ast$ in the timestamp column.
	Apart from the mandatory attributes, the following two custom attributes are present in $L$: The customer $o_1$ has an attribute \texttt{city}, indicating the city of residence of the customer.
	After providing the city initially in 2016, the attribute was updated in 2018, as the customer moved their residence.  
	Moreover, the \texttt{payment reminder} event $e_5$ has an attribute \texttt{fee}, indicating the additional fine incurred by the late payment (e.g., 15\euro).

	The graph in \hyperref[tab:example-oced-objects]{\autoref*{tab:example-oced-general}(a)} shows all objects in $O$ as nodes.
	Edges between nodes indicate the object-to-object (O2O) relationships, that are formally expressed through the \texttt{objects} attributes.
	In the graph of \hyperref[tab:example-oced-events]{\autoref*{tab:example-oced-general}(b)}, events and event-to-object (E2O) relationships are included additionally.
	For readability, the qualifiers for the O2O and E2O relationships are omitted.

For an OCED $L=(E,O,\mi{eaval},\mi{oaval})$, we introduce these shorthands:
\begin{itemize}[nosep]
	\item $E_L = E$ and $O_L = O$ for the set of objects or events of $L$, respectively.
	\item Function $type_L \in \left(O \cup E\right) \rightarrow \left(\objTypeUniv \cup \evTypeUniv\right)$, which assigns \emph{object or event types} to objects or events, defined as:\\
	$$
	type_L(x) = \begin{cases} \mi{oaval}_x(\texttt{type}), &\text{if $x \in O$} \\
		\mi{eaval}_x(\texttt{activity}), &\text{if $x \in E$} \\
	 \end{cases}
	 $$
	\item Function $time_L \in E \rightarrow \mathbb{T}$ with $time_L(e) =\mi{eaval}_e(\texttt{time})$, which maps an event to its timestamp.
	\item Given an optional qualifier $q\in \qualUniv \cup \{\ast\}$, we define the function $obj^q_L\colon (E\cup O) \rightarrow \mathcal{P}(O)$, which assigns an event or object to its set of object references:
	$$
		obj^q_L(x) = \begin{cases}
			\{o \mid (q',o) \in \mi{eaval}_{x}(\texttt{objects}) \land (q = \ast \lor q = q')\}, & \text{if $x \in E$} \\
			\{o \mid (q',o) \in \mi{oaval}_{x}(\texttt{objects}) \land (q = \ast \lor q = q')\}, & \text{if $x \in O$}\end{cases}
		$$
		      For simplicity we also write $obj_L = obj_L^\ast$ for all object references without considering their qualifiers.
\end{itemize}

\section{Object-Centric Querying and Constraints}
\label{sec:querying}
In this section, we detail our approach to object-centric querying and constraints.
First, we introduce the concept of \emph{variable bindings}.
They make up the output of our querying approach.
A variable binding is a collection of concrete objects and events of an OCED that are referred to using variable names.
Through these names, events or objects can be differentiated even if they are of the same type. 

\begin{definition}\label{def:binding}
	Let $L$ be an OCED. The set of variable bindings $\mathbb{B}_L$ under L is:
	\begin{align*}
	\mathbb{B}_L = \{  b_1 \cup b_2 \mid  b_1 \in (\nodeNameUniv \not \rightarrow E_L) \land b_2 \in (\varNamesUniv \not \rightarrow O_L)\}
	\end{align*}
\end{definition}
Consider an example OCED $L$ with $o_1,o_2,o_3 \in O_L$ and $e_1,e_2,e_3 \in E_L$, and let $\texttt{o1}, \texttt{o2}, \texttt{o3} \in \varNamesUniv$ and $\texttt{e1}, \texttt{e2}, \texttt{e3} \in \nodeNameUniv$ be object and event variable names.
Then the following example bindings are part of $\mathbb{B}_L$: $b_1 = \{\}$, $b_2 = \{\texttt{o1} \mapsto o_1 \}$, $b_3 = \{\texttt{o2} \mapsto o_1 \}$, and $b_4 = \{ \texttt{e1} \mapsto e_1, \texttt{e2} \mapsto e_3, \texttt{o1} \mapsto o_1, \texttt{o3} \mapsto o_3 \}$.

In the context of an OCED, we next introduce the concept of child and parent bindings.
A child binding contains all the event and object variables of the parent, and also maps them to the same values as the parent.
\begin{definition}\label{def:parent-child-bindings}
	Let $L$ be an OCED.
	We define the parent-child relation $\sqsubseteq_L$ as follows: For two bindings $p,c \in \mathbb{B}_L$, $p \sqsubseteq_L c \Leftrightarrow  \forall_{x \in \dom(p)}\; p(x) = c(x)$.
	When $p \sqsubseteq_L c$, we call  $c$ a \emph{child binding} of $p$ and $p$ a \emph{parent binding} of $c$.
	Clearly, $\sqsubseteq_L$ is a partial order (i.e., reflexive, antisymmetric and transitive).
\end{definition}
For every OCED $L$, the empty binding $\{\}$ is the smallest element in $\mathbb{B}_L$ regarding $\sqsubseteq_L$.
Considering the previous example bindings, it holds that $b_2 \sqsubseteq_L b_4$ and $b_3 \not\sqsubseteq_L b_4$.
The $\sqsubseteq_L$ relation is useful for describing the output of \emph{nested} queries.

Next, we introduce \emph{binding predicates}.
They allow specifying which bindings a query should return, similar to a \texttt{WHERE} clause in SQL.

\newcommand{\basicFilters}{\textbf{\textsc{BASIC}}_\mathbf{L}}
\begin{definition}\label{def:filter-statement}
	Let $L$ be an OCED.
	Given $L$, a \emph{binding predicate} describes a set of bindings that satisfy this predicate.
	However, different predicates can be distinguished even if they induce the same set of satisfied bindings. 
	We write $\predSet$ for the set of all binding predicates.
	If a binding $b \in \mathbb{B}_L$ satisfies a predicate $s \in \predSet$, we write $b \models s$.
	Additionally, we use the same notation for a set of predicates $S \subseteq \predSet$: $b \models S \Leftrightarrow \forall_{s \in S}\;b \models s$.
\end{definition}
As an example binding predicate for an OCED $L$, consider $s \in \predSet$ with $s \models b \Leftrightarrow b(\texttt{o1}) \in O_L \land b(\texttt{e1}) \in E_L \land b(\texttt{o1}) \in obj_L(b(\texttt{e1}))$ for all $b \in \mathbb{B}_L$.
The binding predicate $s$ encompasses all variable bindings where the variables \texttt{o1} and \texttt{e1} are bound to objects or events of $L$, respectively, such that the values are in an event-to-object relationship.

One can specify (pairwise disjoint) \emph{collections of binding predicates} for different purposes.
Initially, we define the collection $\basicFilters \subseteq \predSet$ in the context of $L$.
Note that our approach is easily extendable by adding more predicate types and collections.
However, for brevity, we do not define further predicates (e.g., based on general data attributes, like the \texttt{price} of an \texttt{orders} object) here.

$\basicFilters$ is made up of three predicate types:
\begin{itemize}[itemsep=2pt,nolistsep,topsep=2pt] \small
	\item \textbf{Event-to-Object} For an event variable $v \in \nodeNameUniv$, an object variable $v' \in \varNamesUniv$ and an optional relationship qualifier $q \in \qualUniv \cup \{  \ast \}$: $ \mathrm{E2O}(v,v',q) \in \basicFilters$, with for any $b \in \mathbb{B}_L$:
	\vspace*{-0.25cm}
	\begin{align*}
		b \models \mathrm{E2O}(v,v',q) \;\Leftrightarrow\; b(v) \in E_L \land b(v') \in O_L \land b(v') \in obj_L^q(b(v)) 
	\end{align*}
	\vspace*{-0.6cm}
	\item \textbf{Object-to-Object} For two object variables $v , v' \in \varNamesUniv$, and an optional qualifier $q \in \qualUniv \cup \{  \ast \}$: $\mathrm{O2O}(v,v',q)\in \basicFilters$, with for any $b\in \mathbb{B}_L$:
	\vspace*{-0.25cm}
	\begin{align*}
		b \models \mathrm{O2O}(v,v',q) \;\Leftrightarrow\; b(v) \in O_L \land b(v') \in O_L \land b(v') \in obj_L^q(b(v)) 
	\end{align*}
	\vspace*{-0.6cm}
	\item \textbf{Time Between Events} For two event variables $v, v' \in \nodeNameUniv$ and a duration interval $t_{min}, t_{max} \in\mathbb{T}$: 	$\mathrm{TBE}(v,v',t_{min},t_{max})  \in \basicFilters$, with for any $b\in \mathbb{B}_L$:
	\vspace*{-0.25cm}
	\begin{align*}
		b \models \mathrm{TBE}(v,v',t_{min},t_{max}) \;\Leftrightarrow\; \,&b(v) \in E_L \land b(v') \in E_L \\
		&\hspace*{0.15cm}\land t_{min} \leq time_L(b(v')) - time_L(b(v))\leq t_{max}
	\end{align*}
	\vspace*{-0.6cm}
\end{itemize}
	Consider the predicate $s_1 = \mathrm{O2O}(\texttt{o1},\texttt{o2},\ast)\in \basicFilters$ and the bindings $b_5 = \{ \texttt{o1} \mapsto o_1 \}$, and $b_6 = \{ \texttt{o1} \mapsto o_1, \texttt{o2}  \mapsto o_2, \texttt{e1} \mapsto e_1 \}$.
	As $b_5$ does not assign $\texttt{o2}$, it holds that $b_5 \not \models s_1$.
	Assuming an object-to-object relation between $o_1$ and $o_2$ exists in $L$,  $b_6 \models s_1$ would hold.

Next, we introduce the concept of \emph{binding boxes}, which correspond to simple queries, yielding sets of variable bindings as output.
\begin{definition}\label{def:binding-box}
	Let $L$ be an OCED.
	A binding box $\mathfrak{b}_L=(\mathrm{Var},\mathrm{Pred})$ over $L$ is a tuple consisting of:
	\begin{itemize}[itemsep=0pt, parsep=0pt,topsep=3pt]
		\item $\mathrm{Var} \in \left\{ ev \cup ob \mid  ev \in  \evVarUniv \not \rightarrow \mathcal{P}(\evTypeUniv) \land ob \in \objVarUniv \not \rightarrow \mathcal{P}(\objTypeUniv) \right\}$, a partial function which specifies to values of which event or object types selected variables should be bound.
		\item $\mathrm{Pred} \subseteq \predSet$, a set of binding predicates.
	\end{itemize}
	Intuitively, $\mathfrak{b}_L$ binds the variable names $\dom(\mathrm{Var})$ to all combination of values (i.e., events or objects of $L$) such that they are of the specified types and the predicate set $S$ holds.
	For convenience, we sometimes write $\mathrm{Var}(\mathfrak{b}_L) = \mathrm{Var}$ and $\mathrm{Pred}(\mathfrak{b}_L) = \mathrm{Pred}$.
	Additionally, we write $\mathfrak{BOX}_L$ for the set of all binding boxes under $L$.
	We define when a binding $b \in \mathbb{B}_L$ satisfies the binding box, written as $b \models \mathfrak{b}_L$, as follows:
	\begin{align*}
		b \models \mathfrak{b}_L \Longleftrightarrow \;   b \models \mathrm{Pred} &\land \dom(b) = \dom(\mathrm{Var}) \\
		& \land \forall_{v \in \dom(\mathrm{Var})}\; \big(b(v) \in E_L \cup O_L \land type_L(b(v)) \in \mathrm{Var}(v)\big)
\end{align*}
\end{definition}
Next, we present a simple example binding box involving one event variable and one object variable.
	For this and further examples, consider an OCED $L=(E,O,\mi{eaval},\mi{oaval})$ of an order management process that is not fully specified here for brevity.
	Consider the simple binding box $\mathfrak{a}_L=(\mathrm{Var},\mathrm{Pred})$, with:
	\begin{itemize}[nosep]
		\item $\mathrm{Var} = \{\texttt{e1} \mapsto \{\texttt{place order}, \texttt{confirm order}\}, \texttt{o1} \mapsto \{\texttt{orders}\}\}$
		\item $\mathrm{Pred} = \left\{ \mathrm{O2E}(\texttt{e1},\texttt{o2},\texttt{order}) \right\}$
	\end{itemize}
For convenience, we will use a visual notation schema for further examples:
Given a binding box $\mathfrak{a}_L = (\mathrm{Var},\mathrm{Pred})$, the elements of $\mathrm{Var}$ are split into multiple lines on the top.
On the bottom (i.e., in the predicate; below the line), the filter predicates $\mathrm{Pred}$ are listed using their representation (e.g., $\mathrm{E2O}$).
	\vspace{-0.33cm}
	\begin{schema}{\mathfrak{a}_{L}}
		\texttt{o1}: \textsc{Object} (\texttt{orders}) \\
		\texttt{e1}: \textsc{Event} (\texttt{place order},\texttt{confirm order})
		\where
		\mathrm{E2O}(\texttt{e1},\texttt{o1},\texttt{order})
	\end{schema}

Given this example binding box  $\mathfrak{a}_L$, we can also construct its output set. 
	Assume that $L$ contains only the objects $o_1,o_2,o_3$ of type \texttt{orders}, where $o_1$ and $o_2$ are associated with a \texttt{place order} event (i.e., with $e_1$ and $e_2$, respectively).
	Additionally, assume that $o_1$ is the only object that is also associated with a \texttt{confirm order} event $e_3$.
	Then we can construct the output bindings of $\mathfrak{a}_L$, $\{ b \in \mathbb{B}_L \mid b \models \mathfrak{a}_L \}$, as: $	out_L(\mathfrak{a}_L) = \big\{\{ \texttt{o1} \mapsto o_1, \texttt{e1} \mapsto e_1\}, \{ \texttt{o1} \mapsto o_2, \texttt{e1} \mapsto e_2 \}, \{ \texttt{o1} \mapsto o_1, \texttt{e1} \mapsto e_3\} \big\}$.

To facilitate nested queries, we define a relation $\preceq_L$ between binding boxes over an OCED $L$.
It encompasses the concept of \emph{refined} binding boxes, where new object or event variables can be introduced, and the filter is at least as strict as before.
\begin{definition}\label{def:leq-between-binding-boxes}
	Let $L$ be an OCED.
	Let $\mathfrak{a}_L, \mathfrak{b}_L \in \mathfrak{BOX}_L$.
	We say $\mathfrak{a}_L \preceq_L \mathfrak{b}_L$ holds if:
		 $\mathrm{Var}(\mathfrak{a}_L) \subseteq \mathrm{Var}(\mathfrak{b}_L)$ and
		 $\mathrm{Pred}(\mathfrak{a}_L) \subseteq \mathrm{Pred}(\mathfrak{b}_L)$.
\end{definition}
For instance, with $\mathfrak{a}_L$ from before and $\mathfrak{b}_L = (\{ \texttt{o1} \rightarrow \{ \texttt{orders}\}\},\emptyset)$, it holds that $\mathfrak{b}_L \preceq_L \mathfrak{a}_L$.
The concept of refined binding boxes enables nested queries, where the first binding box only queries a subset of the overall involved objects or events (e.g., only an \texttt{orders} object but no events).

Next, we define the restriction of binding boxes on only a subset of 
considered predicates.
They enable ignoring certain predicate types for the $\preceq_L$ relation.

\begin{definition}
	Let $L$ be an OCED, let $\mathfrak{a} = (\mathrm{Var},\mathrm{Pred}) \in \mathfrak{BOX}_L$ be a binding box over $L$ and let $X \subseteq \predSet$ be a set of binding predicates.
	The filter-restriction of $\mathfrak{a}$ to $X$, denoted as $\mathfrak{a}\vert_{X}$, is the binding box  $\mathfrak{a}\vert_{X} = (\mathrm{Var},\mathrm{Pred}\cap X)$ over $L$.
\end{definition}
These concepts allow defining a tree structure of binding boxes, where children are refined versions of their parents when considering only basic predicates.
These \emph{query trees} are the core of our approach and enable declarative, nested querying of objects and events.
\newcommand{\boxFunc}{\ensuremath{box}}
\begin{definition}\label{def:binding-box-tree}
	Let $L$ be an OCED.
	A query tree is a tuple $T=(V,F,r,l,\boxFunc)$:
	\begin{itemize}[nosep]
		\item $V$ is a finite set of nodes.
		\item $r \in V$ is the designated root node (i.e., the only node with no parent).
		\item $F \subseteq V \times V$ is a set of edges between nodes, such that in the directed graph $(V,F)$ there is exactly one path from the root $r$ to $a$ for all $a \in V$.
		\item $l \colon F \rightarrow \bindingSetNames$ is an injective function, assigning unique names to edges.
		\item $\boxFunc\colon V \rightarrow \mathfrak{BOX}_L$ is a function which maps each node in $V$ to a binding box over $L$, such that for all edges $(a,b) \in F$ with $\boxFunc(a) = \mathfrak{a}$ and $\boxFunc(b) = \mathfrak{b}$, it holds that $\mathfrak{a}\vert_{\basicFilters} \preceq_L \mathfrak{b}\vert_{\basicFilters}$.
	\end{itemize}
\end{definition}

	Next, in \autoref{fig:binding-tree-example}, we show an example query tree only using predicates from $\basicFilters$.
	Afterwards, we introduce more complex queries and constraint examples which motivate the specified restriction of the $\boxFunc$ function.

	\definecolor{caribbeangreen}{rgb}{0.0, 0.8, 0.29}
	\begin{figure}[h]
	\begin{minipage}[c]{0.5\linewidth}
		\scriptsize
		\begin{schema}{\boxFunc(v_{0})}
			\texttt{o1}: \textsc{Object} (\texttt{orders})\\ \texttt{e1}: \textsc{Event} (\texttt{confirm order})\\
			\where \mathrm{E2O}(\texttt{e1},\texttt{o1},\texttt{$\ast$})
		\end{schema}
		\vspace{-0.9cm}
		\begin{schema}{\boxFunc(v_{1})}
			{\color{darkgray} \texttt{o1}: \textsc{Object} (\texttt{orders})}\\
			{\color{darkgray} \texttt{e1}: \textsc{Event} (\texttt{confirm order})}\\
			\texttt{e2}: \textsc{Event} (\texttt{pay order})\\
			\where
			{\color{darkgray} \mathrm{E2O}(\texttt{e1},\texttt{o1},\texttt{$\ast$})} \\
			\mathrm{E2O}(\texttt{e2},\texttt{o1},\texttt{$\ast$})\\
			\mathrm{TBE}(\texttt{e1},\texttt{e2},\text{0},\text{4w})
		\end{schema}
		\vspace{-0.9cm}
		\begin{schema}{\boxFunc(v_{2})}
			{\color{darkgray} \texttt{o1}: \textsc{Object} (\texttt{orders})}\\
			{\color{darkgray} \texttt{e1}: \textsc{Event} (\texttt{confirm order})}\\
			\texttt{e2}: \textsc{Event} (\texttt{payment reminder})\\
			\where
			{\color{darkgray} \mathrm{E2O}(\texttt{e1},\texttt{o1},\texttt{$\ast$})} \\
			\mathrm{E2O}(\texttt{e2},\texttt{o1},\texttt{$\ast$})
		\end{schema}
		\vspace{-0.6cm}
	\end{minipage}\hfill\begin{minipage}[c]{0.5\linewidth}
		\begin{figure}[H]
			\renewcommand{\arraystretch}{0.8}
			\centering
			\begin{tikzpicture}
				\node (-1) at (0, 1.8) {};
				\node (0) at (0, 1.2) {$v_0$};
				\node (1) at (-0.8, 0.25) {$v_1$};
				\node (2) at (0.8, 0.25) {$v_2$};
				\draw [->] (-1) -- (0);
				\draw [->] (0) -- (1) node [midway, xshift=-7, yshift=5] {$\texttt{A}$};
				\draw [->] (0) -- (2) node [midway, xshift=7, yshift=5] {$\texttt{B}$};
				\node[above right=-1.5cm and 1.4cm of 0] (t0) {\footnotesize
					\begin{tabular}[t]{c|c}
						\texttt{o1}           & \texttt{e1}           \\ \hline
						\color{cyan} $o_1$    & \color{cyan}  $e_1$   \\
						\color{magenta} $o_2$ & \color{magenta} $e_2$ \\
						\color{caribbeangreen} $o_3$    & \color{caribbeangreen} $e_3$    \\
						\color{orange} $o_4$  & \color{orange} $e_4$  \\
					\end{tabular}
				};
				\draw[dashed,color=gray] (0) -- (t0);
				\node[below right=8pt and -0.2cm of 1] (t1) {\footnotesize
					\begin{tabular}[t]{c|c|c}
						\color{gray} \texttt{o1} & \color{gray} \texttt{e1} & \texttt{e2}        \\ \hline
						\color{caribbeangreen} $o_3$       & \color{caribbeangreen} $e_3$       & \color{caribbeangreen} $e_7$ \\
						\color{caribbeangreen} $o_3$       & \color{caribbeangreen} $e_3$       & \color{caribbeangreen} $e_8$ \\
					\end{tabular}
				};
				\draw[dashed,color=gray] (1) -- (t1);
				\node[below right=4pt and 0.4cm of 2] (t2) {\footnotesize
					\begin{tabular}[t]{c|c|c}
						\color{gray} \texttt{o1} & \color{gray}\texttt{e1} & \texttt{e2}           \\ \hline
						\color{cyan} $o_1$       & \color{cyan}  $e_1$     & \color{cyan} $e_5$    \\
						\color{magenta} $o_2$    & \color{magenta} $e_2$   & \color{magenta} $e_6$ \\
						\color{caribbeangreen} $o_3$       & \color{caribbeangreen} $e_3$      & \color{caribbeangreen} $e_9$    \\
					\end{tabular}
				};
				\draw[dashed,color=gray] (2) -- (t2);
			\end{tikzpicture}
			{\scriptsize
			\vspace*{-0.45cm}
			\begin{schema}{\Delta\boxFunc(v_{1})}
				\texttt{e2}: \textsc{Event} (\texttt{pay order})\\
				\where
				\mathrm{E2O}(\texttt{e2},\texttt{o1},\texttt{$\ast$})\\
				\mathrm{TBE}(\texttt{e1},\texttt{e2},\text{0},\text{4w})
			\end{schema}
			\vspace{-0.9cm}
			\begin{schema}{\Delta\boxFunc(v_{2})}
				\texttt{e2}: \textsc{Event} (\texttt{payment reminder})\\
				\where
				\mathrm{E2O}(\texttt{e2},\texttt{o1},\texttt{$\ast$})
			\end{schema}
			\vspace{-0.5cm}}
		\end{figure}
	\end{minipage}
	\caption{A query tree with three nodes.
	On the left, the binding box of each node is shown.
	On the top right, the tree structure is visualized with example output tables.
	The boxes on the bottom right, marked with $\Delta$, only show additions to their parents.
	}
	\label{fig:binding-tree-example}
\end{figure}

In \autoref{fig:binding-tree-example}, the query tree $T_1 = (V,F,r,l,\boxFunc)_L$ is shown, with $V=\{v_0,v_1,v_2\}$, $r = v_0$, $F = \{(v_0,v_1),(v_0,v_2)\}$, $l((v_0,v_1)) = \texttt{A}$, and $l((v_0,v_2)) = \texttt{B}$.
	The top right of \autoref{fig:binding-tree-example} shows the graph of $T_1$ with exemplary output binding tables, while $\boxFunc$ is presented on the left.
	Naturally, the child binding boxes contain many duplicates (shown in gray).
	To ease readability, we omit variables and predicates that are already present in the binding box of the parent node in future examples.
	With these omissions (marked using $\Delta$), $\boxFunc(v_1)$ and $\boxFunc(v_2)$ can be presented more compactly, as shown on the bottom right.

	On the top right of \autoref{fig:binding-tree-example}, we show exemplary output sets of $\boxFunc(v_0)$, $\boxFunc(v_1)$, and $\boxFunc(v_2)$ as tables next to the corresponding nodes.
	The rows for $v_0$ are colored in four different colors.
	For $v_1$ and $v_2$, each output row is colored based on $\sqsubseteq_L$, indicating from which parent binding in the output set of $v_0$ the row is derived.
	The first two output binding rows for $v_0$ (in \textcolor{cyan}{cyan} and \textcolor{magenta}{magenta})  have exactly one child binding in the output set of $v_1$ and none in the output set of $v_2$.
	For the third output binding row of $v_0$ (in \textcolor{caribbeangreen}{green}), one child binding in $v_1$ exists, and there are also two child bindings in the output set of $v_2$.
	The last output row of $v_0$ (in \textcolor{orange}{orange}) has no child binding in the output sets of $v_1$ or $v_2$.
\newcommandx{\setFilters}[2][1=u,2=T]{\textbf{\textsc{CHILD SET}}^{#2}_{#1}}

In a nested query, oftentimes, the result of the inner query is used in the outer query in an aggregated way.
For instance, in a query for all customers with more than 100 orders, the outer query (all customers) uses the result count of the inner query (all orders by the customer) as its filter.
To express such queries, we introduce a new set of binding predicates, $\setFilters$, in the context of a query tree $T=(V,F,r,l,\boxFunc)$ and one of its nodes $u \in V$.
For every child node $v \in V$ with $(u,v) \in F$ and $l((u,v)) = A$, predicates of the form $\mathrm{CBS}(A,n_{min},n_{max})$ with $n_{min},n_{max} \in \mathbb{N}_0$ are available in $\setFilters$.
They are fulfilled for a binding $b\in \mathbb{B}_L$, when the set of child bindings of $b$ in $v$, $S = \{ x \in \mathbb{B}_L \mid x \models \boxFunc(v) \land b \sqsubseteq_L x \}$, is in the specified size range (i.e., $n_{min} \leq \left|S\right| \leq n_{max}$).
Note, that these predicates can be recursive, as the predicates of a child node are already considered when evaluating the predicates of the parent node.
In particular, they are also only well-defined for the binding box of a specified node, which is why they are not considered for the $\preceq_L$ relation in the tree, which only considers the predicates in $\basicFilters$.

\autoref{fig:example-set-filters} shows such an extended version of the previous tree example.

\begin{figure}[h]
	\begin{minipage}[c]{0.55\linewidth}\footnotesize
	\begin{schema}{\boxFunc(v_0)}
		\texttt{o1}:  \textsc{Object} (\texttt{orders}) \\
		\texttt{e1}: \textsc{Event} (\texttt{confirm order})
		\where
		\mathrm{E2O}(\texttt{e1},\texttt{o1},\ast) \\
		\mathrm{CBS}(\texttt{A},0,0)  \text{\color{gray}\tiny \hspace{0.5cm} //  $\in \setFilters[v_0][T_2]$ } \\
		\mathrm{CBS}(\texttt{B},0,0)  \text{\color{gray}\tiny \hspace{0.5cm} //  $\in \setFilters[v_0][T_2]$ } \\
	\end{schema}
\end{minipage}\hfill\begin{minipage}[c]{0.45\linewidth}
	\begin{figure}[H]
		\renewcommand{\arraystretch}{0.8}

		\centering
		\begin{tikzpicture}
			\node (-1) at (0, 1.5) {};
			\node (0) at (0, 1.0) {$v_0$};
			\node (1) at (-0.7, 0.25) {$v_1$};
			\node (2) at (0.7, 0.25) {$v_2$};
			\draw [->] (-1) -- (0);
			\draw [->] (0) -- (1) node [midway, xshift=-7, yshift=5] {$\texttt{A}$};
			\draw [->] (0) -- (2) node [midway, xshift=7, yshift=5] {$\texttt{B}$};
			\node[below right=0.35cm and -0.8cm of 0] (t0) {\footnotesize
				\begin{tabular}[t]{c|c}
					\texttt{o1}                & \texttt{e1}               \\ \hline
					\color{cyan} \st{$o_1$}    & \color{cyan} \st{$e_1$}   \\
					\color{magenta} \st{$o_2$} & \color{magenta}\st{$e_2$} \\
					\color{caribbeangreen}\st{$o_3$}     & \color{caribbeangreen}\st{$e_3$}    \\
					\color{orange} $o_4$       & \color{orange} $e_4$      \\
				\end{tabular}
			};
			\draw[dashed,color=gray] (0) -- (t0);
			\node[below left=0pt and 0.2cm of 1] (t1) {\footnotesize
				\begin{tabular}[t]{c|c|c}
					\color{gray} \texttt{o1} & \color{gray} \texttt{e1} & \texttt{e2}        \\ \hline
					 \color{caribbeangreen} $o_3$       & \color{caribbeangreen} $e_3$       & \color{caribbeangreen} $e_7$ \\
					\color{caribbeangreen} $o_3$       & \color{caribbeangreen} $e_3$       & \color{caribbeangreen} $e_8$ \\
				\end{tabular}
			};
			\draw[dashed,color=gray] (1) -- (t1);
			\node[below right=-5pt and 0.2cm of 2] (t2) {\footnotesize
				\begin{tabular}[t]{c|c|c}
					\color{gray} \texttt{o1} & \color{gray}\texttt{e1} & \texttt{e2}           \\ \hline
					\color{cyan} $o_1$       & \color{cyan}  $e_1$     & \color{cyan} $e_5$    \\
					\color{magenta} $o_2$    & \color{magenta} $e_2$   & \color{magenta} $e_6$ \\
					\color{caribbeangreen} $o_3$       & \color{caribbeangreen} $e_3$      & \color{caribbeangreen} $e_9$    \\
				\end{tabular}
			};
			\draw[dashed,color=gray] (2) -- (t2);
		\end{tikzpicture}
	\end{figure}
\end{minipage}
\caption{An extension of the query tree from \autoref{fig:binding-tree-example} with child filter predicates ($\mathrm{CBS}$).
}
\label{fig:example-set-filters}
\end{figure}

\autoref{fig:example-set-filters} shows a new query tree $T_2=(V,F,r,l,\boxFunc)$, with $V=\{v_0,v_1,v_2\}$, $r = v_0$, and $F = \{(v_0,v_1),(v_0,v_2)\}$.
	While $v_1$ and $v_2$ are the same as in \autoref{fig:binding-tree-example}, the root node box ($v_0$) now contains two additional predicates of type $\setFilters[v_0][T]$.
	$T_2$ queries placed orders that \emph{were not paid fast} (i.e., within 4 weeks after confirmation) and for which also \emph{no payment reminder was sent}.
	Again, we annotate example output tables for each node in the tree on the right of \autoref{fig:example-set-filters}.
If bindings are removed only by a child set predicate of a binding box but fulfill the basic predicates of it (i.e., $\basicFilters$), we indicate this by including this binding with a strikethrough.
Generally, children of the node might still contain child bindings for crossed out parent binding rows (e.g., all rows for $v_1$ and $v_2$ in \autoref{fig:example-set-filters}); however, in practice, they are often not of particular interest and thus sometimes ignored or omitted from the result tables.

Next, we want to outline how the presented querying approach can be extended.
In general, the output tables of the query tree nodes can be augmented and filtered freely.
As augmentation, labels for each output row can be computed and added as columns (e.g., the total order volume of a customer).
In the following, we will describe how \emph{constraints} can be implemented.
At its core, constraints are a special case of such general labels, defining for each output row if it should be considered \emph{satisfied} or \emph{violated}.
As this classification is binary, a set of predicates for each node can be used for specifying the violation criteria.
For a tree node $v \in V$, we write $constr(v) \subseteq \predSet$ for its set of constraint predicates.
Fully defining these additions formally is outside the scope of this paper.
However, in the following, we present a short example as a demonstration:
Consider the query tree constraint $C=((V,F,r,l,\boxFunc),\constrFuncStandalon)$ shown in \autoref{fig:example-constraint}, with $V=\{v_0,v_1\}$, $r = v_0$ and $F = \{(v_0,v_1)\}$. The graph $(V,F)$ with the edge labels~$l$ is shown on the right, while $\boxFunc$ with $\constrFuncStandalon$ is presented on the left.

\begin{figure}[h]
    \centering
    \begin{minipage}[c]{0.6\linewidth}
        \scriptsize
        \begin{schema}{\boxFunc(v_{0}) \text{ with } \constrFunc{v_{0}}}
            \texttt{o1}: \textsc{Object} (\texttt{orders})\\ \texttt{e1}: \textsc{Event} (\texttt{confirm order})\\
            \where \mathrm{E2O}(\texttt{e1},\texttt{o1},\texttt{$\ast$})
            \where \mathrm{CBS}(\texttt{A},1,1) \hspace{0.5cm} \text{\color{gray} // $\in \constrFunc{v_0}$}
        \end{schema}
        \vspace{-0.85cm}
        \begin{schema}{\Delta\boxFunc(v_{1}) \text{ with } \constrFunc{v_{1}}}
            \texttt{e2}: \textsc{Event} (\texttt{pay order})\\
            \where \mathrm{E2O}(\texttt{e2},\texttt{o1},\texttt{$\ast$})\\
            \mathrm{TBE}(\texttt{e1},\texttt{e2},\text{0},\text{4w})
            \where
        \end{schema}
        \vspace{-0.33cm}

    \end{minipage}\hfill\begin{minipage}[c]{0.39\linewidth}
        \begin{figure}[H]

			\renewcommand{\arraystretch}{0.8}
            \centering
            \begin{tikzpicture}

				\node (-1) at (0, 2.33) {};
				\node (0) at (0, 1.66) {$v_0$};
				\node (1) at (0, 0.25) {$v_1$};
				\draw [->] (-1) -- (0);
				\draw [->] (0) -- (1) node [midway, xshift=-7, yshift=5] {$\texttt{A}$};

                \node[above right=-0.5cm and 1.4cm of 0] (t0) {\footnotesize
                    \begin{tabular}[t]{c|c|c}
                        \texttt{o1}           & \texttt{e1}           & \scriptsize \textbf{\emph{Satisfied}} \\ \hline
                        \color{cyan} $o_1$    & \color{cyan}  $e_1$   & \color{cyan} \checkmark                 \\
                        \color{magenta} $o_2$ & \color{magenta} $e_2$ & \color{magenta}  \checkmark             \\
                        \color{caribbeangreen} $o_3$    & \color{caribbeangreen} $e_3$    & \color{caribbeangreen} $\cross$                 \\
                        \color{orange} $o_4$  & \color{orange} $e_4$  & \color{orange} $\cross$                   \\
                    \end{tabular}
                };
                \draw[dashed,color=gray] (0) -- (t0);
                \node[below right=-1.4cm and 1.4cm of 1] (t1) {\footnotesize
                    \begin{tabular}[t]{c|c|c}
                        \color{gray} \texttt{o1} & \color{gray} \texttt{e1} & \texttt{e2}           \\ \hline
                        \color{cyan} $o_1$       & \color{cyan}  $e_1$      & \color{cyan} $e_7$    \\
                        \color{magenta} $o_2$    & \color{magenta} $e_2$    & \color{magenta} $e_8$ \\
                        \color{caribbeangreen} $o_3$       & \color{caribbeangreen} $e_3$       & \color{caribbeangreen} $e_9$    \\
                        \color{caribbeangreen} $o_3$       & \color{caribbeangreen} $e_3$       & \color{caribbeangreen} $e_{10}$ \\
                    \end{tabular}
                };
                \draw[dashed,color=gray] (1) -- (t1);
            \end{tikzpicture}
        \end{figure}
    \end{minipage}
    \caption{
        Example constraint specifying that every confirmed order \emph{should} be paid within 4 weeks after the confirmation exactly once.
        For each $v \in V$, the set $\constrFunc{v}$ is shown in the corresponding binding box below an extra line.
    In the example output tables shown on the right, this constraint is satisfied for all binding rows of $\boxFunc(v_0)$ except the last two rows.
    The violation status of an output binding of $\boxFunc(v_0)$ is annotated to the corresponding output row as either $\checkmark{}$~or~$\cross{}$.}
    \label{fig:example-constraint}
\end{figure}

\section{Implementation}
\label{sec:impl}
We implemented the querying and constraint checking approach introduced in \autoref{sec:querying} as the full-stack application \emph{OCPQ} consisting of two parts:
 (1) A high-performance execution engine backend implemented in Rust and (2) an interactive user-friendly query editor frontend written in Typescript.
The tool is publicly available at \url{https://github.com/aarkue/OCPQ}, and there are installers as well as further resources and guides available at \url{https://ocpq.aarkue.eu}.

Describing the implementation, and all its features, in detail is outside the scope of this paper.
However, we briefly mention some key aspects:\\
\textbf{Importing \& Pre-processing:} All file types of the OCEL~2.0 specification can be imported.
The OCEDs are then pre-processed to enable fast query execution.\\
\textbf{Representation of Bindings:} Variable bindings are represented in a memory-efficient way, encoding the variable name and its value only through an 8-byte integer each.
This enables faster execution times and usage for larger datasets.\\
\textbf{Parallelized Recursive Query Execution Algorithm:} Queries are executed by a recursive algorithm that allows for full parallelization between bindings. \\
\textbf{HPC Deployment:} To leverage this parallelization, the tool allows users to easily deploy query execution on a High-Performance Computing (HPC) cluster. \\
\textbf{Intelligent Binding Order:} New variables are bound in an efficient order, applying predicate filters as early as possible to remove unwanted bindings.

\begin{figure}[h]
    \centering
    \includegraphics[width=0.93\textwidth]{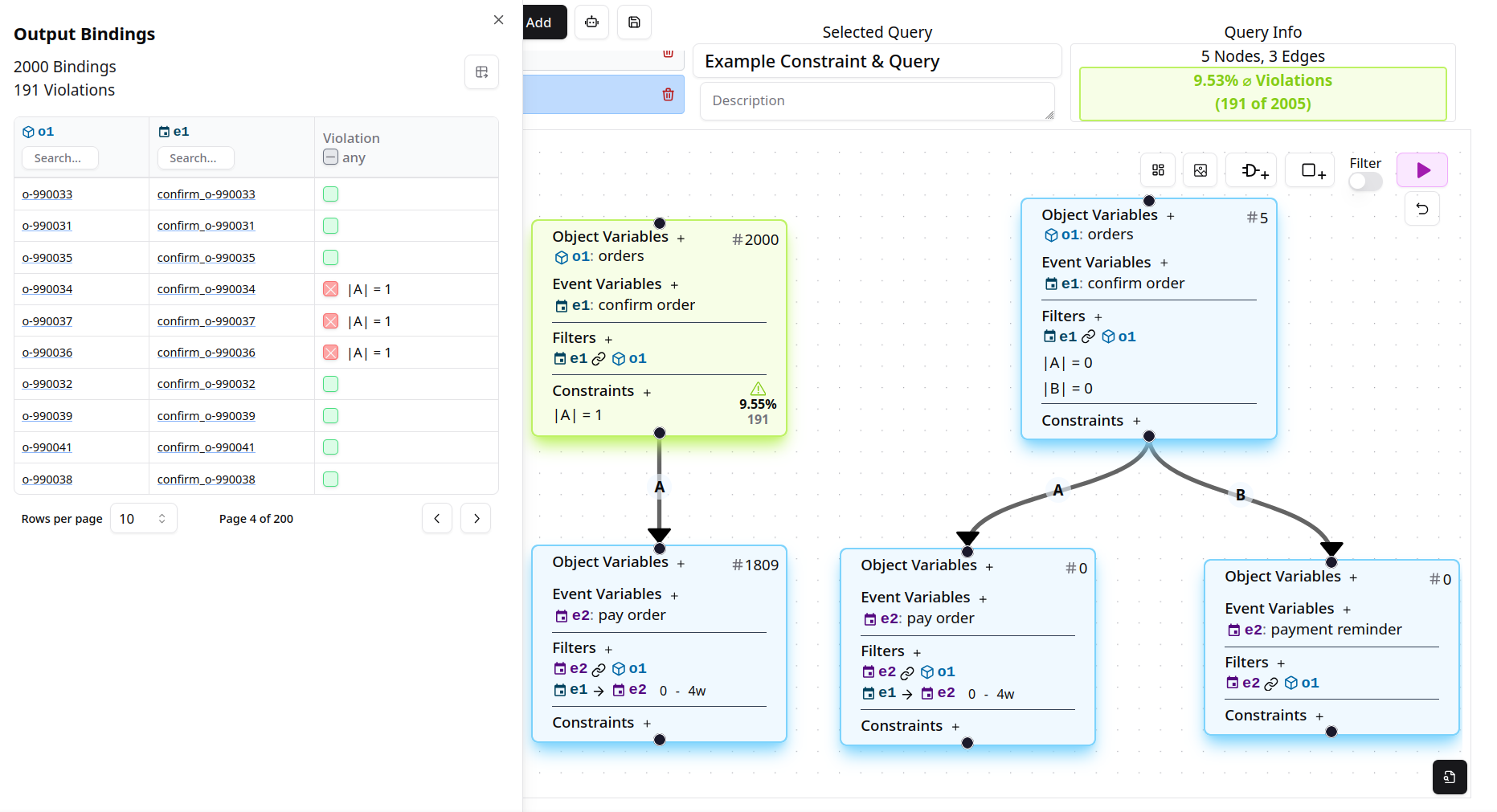}
    \caption{Screenshot of the implemented OCPQ tool.
    The two shown trees correspond to the examples in this paper (i.e., \autoref{fig:example-constraint} on the left, and \autoref{fig:example-set-filters} on the right).
    The root node of the left tree is colored according to the percentage of violations (9.55\%) and its output binding results are shown in the table on the left.}
    \label{fig:implementation-screenshot}
\end{figure}

A screenshot of the tool is shown in \autoref{fig:implementation-screenshot}.
The frontend editor visualizes query trees similar to the notation style that is used throughout this paper.

\section{Evaluation}
\label{sec:eval}
For evaluation, we created seven queries and constraints for a real-life OCED dataset of a loan application process.
The OCED was derived from the BPI Challenge 2017 log~\cite{vandongenBPIChallenge2017}, and has more than 1,200,000 events and 100,000 objects.

The seven example queries (Q1 -- Q7) are designed to be of real-life relevancy and cover a large variety of concepts, ranging from simple to more complex.
In the following, we briefly introduce the used queries, first categorizing their concept or type and then describing the concrete query in natural language:
{
    \scriptsize
    \begin{description}[itemsep=1.5pt,topsep=4pt]\baselineskip=-10000pt\lineskip=0pt
        \item[Q1] A simple constraint on the number of events that should be associated with an object: ``Every application should be submitted exactly once.''
        \item[Q2] A basic eventually-follows constraint (under a specified object type): ``Every Offer should be returned at least once after creation.''
        \item[Q3] A constraint on the number of objects of a type to be associated with events: ``Each O\_Returned event should involve exactly one Offer.''
        \item[Q4] An eventually-follows constraint spanning across events of two related objects: ``After an Application was accepted, there should be at least one associated Offer accepted afterwards.''
        \item[Q5] A constraint enforcing that an object is associated with an event if it was involved with another event: ``The Resource that accepts an Application should also create all Offers for that Application.''
        \item[Q6] A query for the maximal duration between two events, both associated with an object: ``What is the maximum delay between an Offer being created and accepted?''
        \item[Q7] A query for multiple object and event instances of the same type that are linked through another object: ``Get all combinations of two Offers that are associated with the same Application and the corresponding Offer creation events.''
    \end{description}
    }
As qualitative evaluation, to investigate expressiveness, we analyzed for each query whether it (or an equivalent constraint) can be modeled in DECLARE\footnote{As DECLARE is based on traditional, flat event data, we assumed a reasonable flat representation of the OCED on only one object type for evaluating its expressiveness.}~\cite{pesicDECLAREFullSupport2007}, OCCG~\cite{parkMonitoringConstraintsBusiness2022}, or OCBC~\cite{aalstObjectCentricBehavioralConstraints2017}.
As the implementations of these approaches are primarily research prototypes, and are either very slow or completely unusable for larger datasets, we did not measure their execution times.
Instead, as quantitative evaluation, we compared the query execution duration of OCPQ to the general querying platforms Neo4j (used in~\cite{esserMultiDimEventData2021}), SQLite (one of the official OCEL~2.0 formats), and the performance-focused DuckDB~\cite{DuckDB_Raasveldt}, to demonstrate the runtime performance and scalability of OCPQ.

\begin{figure}[h] \centering
    \subfloat[Q4 in OCPQ]
    {
        \includegraphics[width=0.29\textwidth]{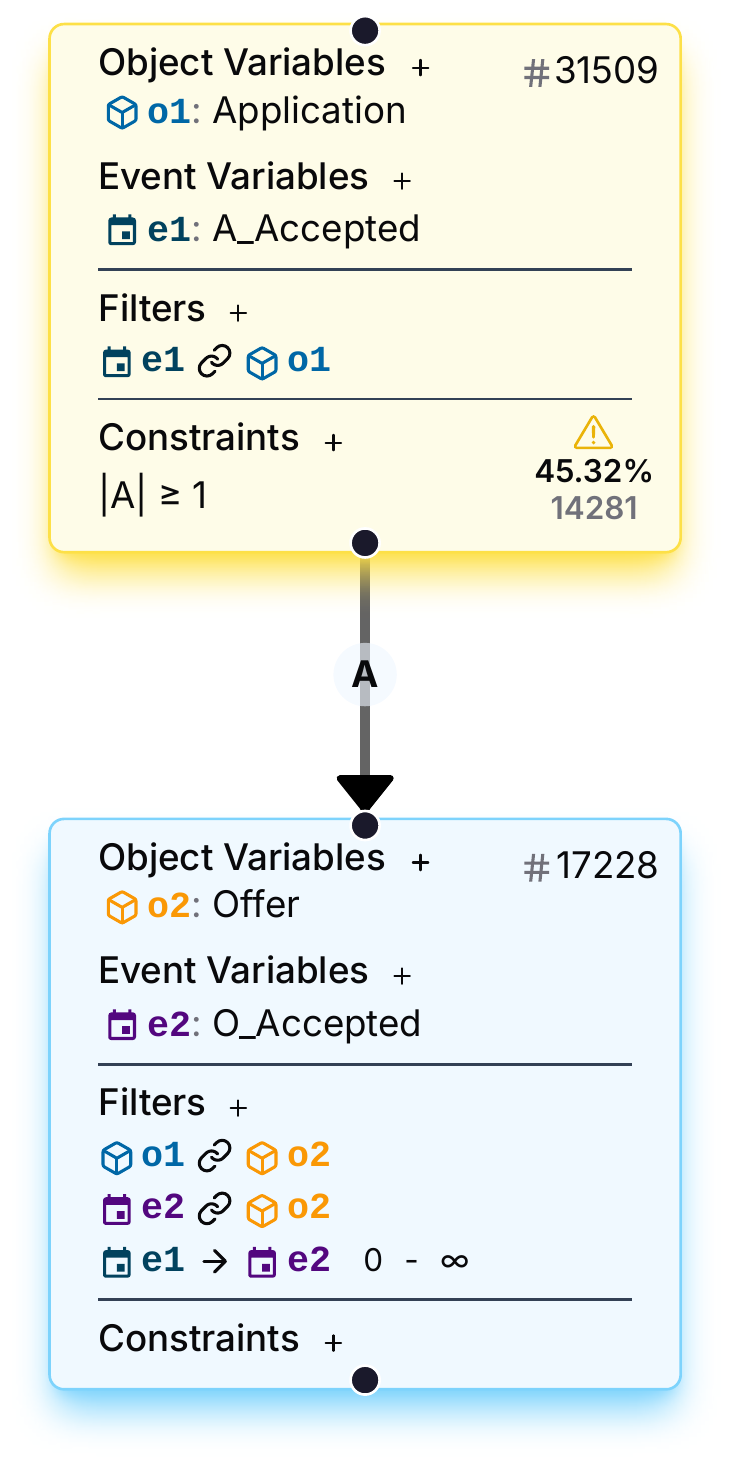}} \hfill
        \subfloat[Q4 in SQL]
        {
    \includegraphics[width=0.68\textwidth]{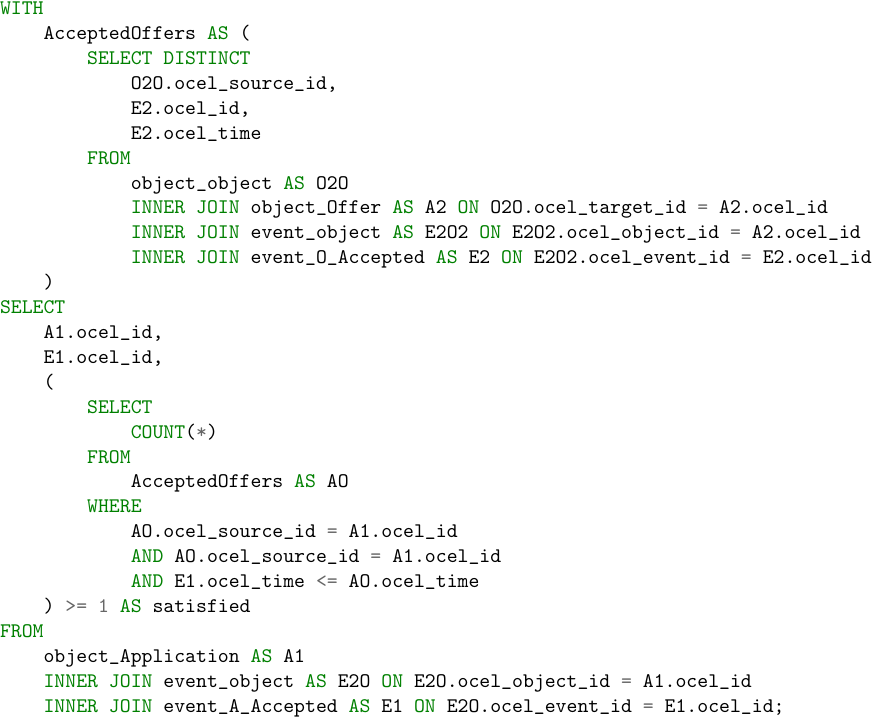}}\\[8pt]
    \subfloat[Q4 in Cypher (violation count only)]
    {
    \includegraphics[width=0.87\textwidth]{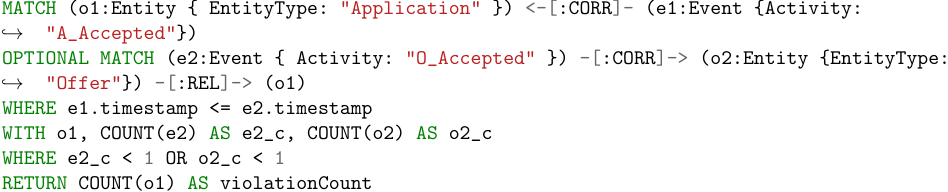}}
    \caption{Constraint Q4 formulated in OCPQ, SQL, and Cypher.}
    \label{fig:q4-formulations-example}
\end{figure}

\autoref{fig:q4-formulations-example} shows how Q4 has been formulated in OCPQ, SQL (for SQLite and DuckDB), and Cypher (for Neo4j), as an example.
While evaluating the usability of OCPQ in detail is outside the scope of this paper, this example still demonstrates the complexity of implementing simple business queries in general querying languages, like SQL or Cypher.

For SQLite, the OCEL~2.0 database was completely loaded into memory, to be a more accurate comparison to OCPQ, and it was ensured that appropriate table indices were added.
For Neo4j, the database dump\footnote{See \url{https://data.4tu.nl/datasets/5c9717a0-4c22-4b78-a3ad-d2234208bfd7/1}.} from \cite{esserMultiDimEventData2021}, was imported in a compatible version of Neo4j (3.5.35) and additionally Neo4j was configured to allow extensive memory usage.
To measure accurate execution times for Neo4j, we used query formulations that report either the total count of results (for queries) or violations (for constraints).
This was done to exclude misleading execution times for when the result rows are being streamed instead of fully computed.
Moreover, these numbers are also always calculated in OCPQ.

We used the full current implementation for formulating the example queries using OCPQ, also including features that were only briefly mentioned in this paper.
All queries except Q6 are modeled fully visually and without programming in OCPQ.
For Q6, a scripting feature of the tool is used to calculate the maximum duration of subquery results as a label. 
More details on the evaluation, including the dataset, raw execution times, and formulations of the queries across all applicable approaches are available at \url{https://github.com/aarkue/ocpq-eval}.

\begin{figure}[h]
    \centering
\includegraphics[trim={0 0 0 2.6cm},clip,width=0.98\textwidth]{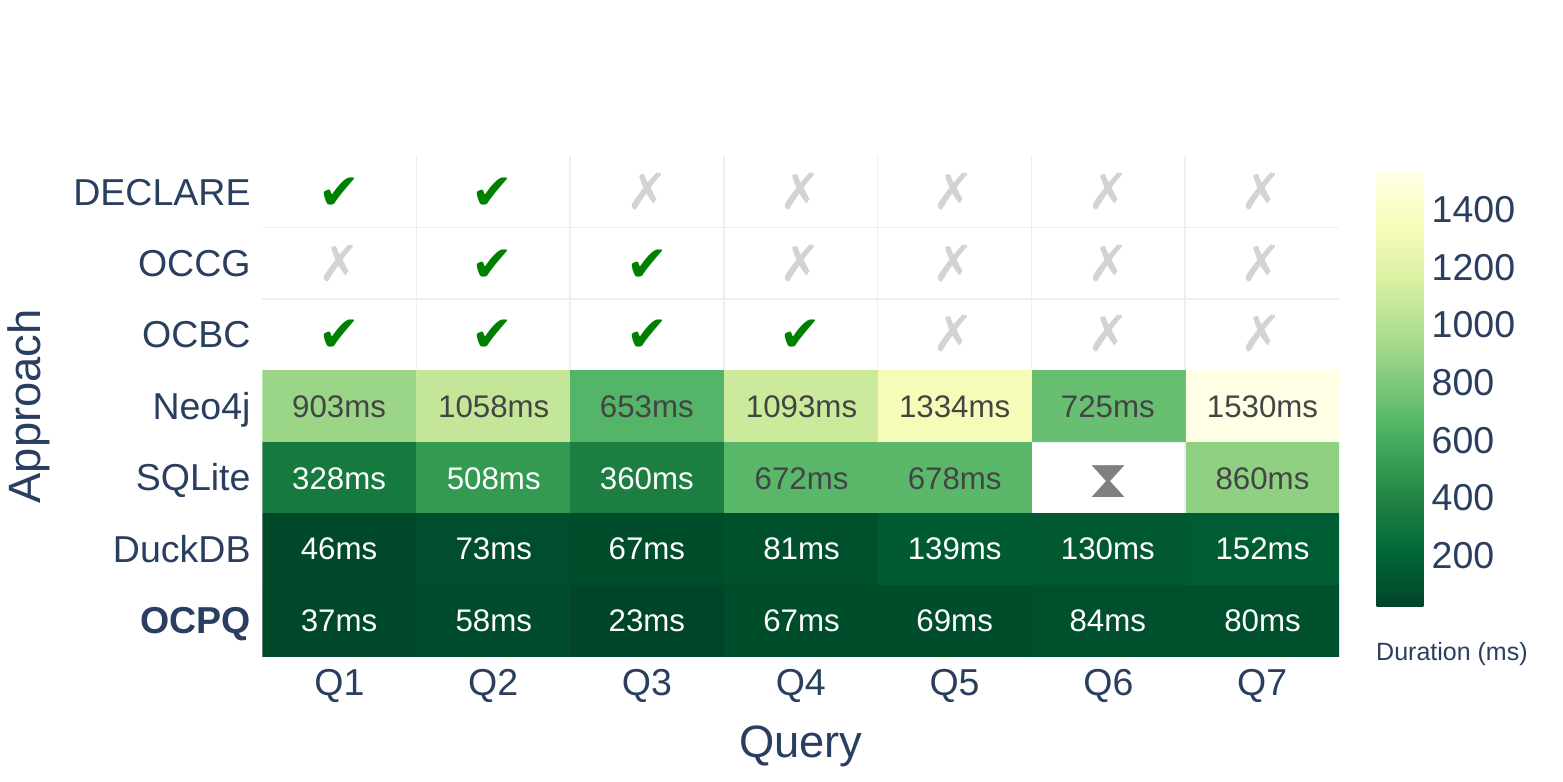}
\caption{Combined qualitative and quantitative evaluation of OCPQ.
For the first three approaches, we only specify if the query/constraint can be expressed or not.
For Neo4j, SQLite, DuckDB, and OCPQ, all queries are expressible, and we report their mean execution time across ten runs on the BPIC2017 OCED.
Evaluating Q6 in SQLite took longer than four minutes and is thus omitted.
}
\label{fig:eval-heatmap}
\end{figure}

The combined results for both evaluations are shown in \autoref{fig:eval-heatmap}.
The results show two things very clearly:
First, there is a large gap in expressiveness, resulting in many relevant queries and constraints that are not representable in previously proposed visual approaches, like DECLARE, OCCG, or OCBC.
In particular, the queries Q5 -- Q7 cannot be modeled in any of them.
Second, also general-purpose querying solutions, like SQLite or Neo4j, are not well suited for important types of OCED queries and are significantly slower than OCPQ for all tested queries.
While DuckDB performs similar to OCPQ in terms of execution times, SQL formulations of OCED queries quickly grow complex and are difficult to write, read, and interpret.
Moreover, a simple SQL translation also does not yield any subquery results, unlike OCPQ where results are available for each subquery. 

Of course, there are some limitations to our evaluation.
As we created the queries Q1 -- Q7 ourselves, they might exhibit certain biases.
However, they still serve as typical examples with real-world relevancy and cover a wide range of constructs.
Also, while we spent special attention to crafting well-performing queries for both SQLite and Neo4j, even more efficient formulations might be possible.
Finally, we performed our evaluation only on one real-life OCED dataset.

\section{Conclusion}
\label{sec:conclusion}
In this paper, we proposed an object-centric querying and constraint approach, \emph{OCPQ}, based on variable bindings of objects and events.
Through the use of bindings, it can express more advanced queries and constraints than previous graphical approaches.
We implemented our approach as a full-stack solution, supporting efficient execution of queries and constraints, as well as an interactive editor for creating them.
As evaluation, we constructed several example queries and constraints and compared OCPQ to other approaches.
The evaluation demonstrated the limitations of previous work in terms of expressiveness, and showed that OCPQ also significantly outperforms several general querying solutions, namely SQLite and Neo4j, in terms of runtime.

In future work, we plan to conduct a detailed performance analysis on more datasets.
Additionally, the expressiveness of the proposed approach has to be studied systematically and in more depth.
Furthermore, we see a lot of potential for extensions.
Our concept of object-centric querying is very universal, allowing applications for \emph{process constraints}, \emph{OCED filtering}, \emph{general annotations}, as well as for generating \emph{situation tables} as input for machine learning techniques.

\begin{credits}
    \subsubsection{\ackname} 
    The authors gratefully acknowledge the German Federal Ministry of Education and Research (BMBF) and the state government of North Rhine-Westphalia for supporting this work as part of the NHR funding.

    \subsubsection{\discintname}
    The authors have no competing interests to declare that are relevant to the content of this paper.
\end{credits}

%
%
\bibliographystyle{splncs04}
\bibliography{references}
\end{document}